\documentclass[10pt,a4paper]{article}
\pdfoutput=1

\usepackage{amsmath}
\numberwithin{equation}{section}
\usepackage{graphicx,subfigure,color}
\usepackage[font=footnotesize,labelfont=bf]{caption}
\begin{document}
\def\thefootnote{\fnsymbol{footnote}}

\begin{center}
\Large{\textbf{A divergence-free parametrization for dynamical dark energy}} \\[0.3cm]

\large{\"{O}zg\"{u}r Akarsu$^{\rm a}$, Tekin Dereli$^{\rm a}$, J. Alberto Vazquez$^{\rm b}$}
\\[0.3cm]

\small{
\textit{$^{\rm a}$ Department of Physics, Ko\c{c} University, 34450 Sar{\i}yer, {\.I}stanbul, Turkey}}

\vspace{.1cm}

\small{
\textit{$^{\rm b}$ Brookhaven National Laboratory (BNL), Department of Physics, Upton, NY 11973-5000, USA}}

\end{center}

\vspace{.4cm}

\hrule \vspace{0.3cm}
\noindent \small{\textbf{Abstract}\\
We introduce a new parametrization for the dark energy, led by the same idea to the 
linear expansion of the equation of state in scale factor $a$ and in redshift $z$, which diverges neither in the past nor future and contains the same number of degrees of freedom with the former two. We present constraints of the cosmological parameters using the most updated baryon acoustic oscillation (BAO) measurements along with cosmic microwave background (CMB) data and a recent reanalysis of Type Ia supernova (SN) data. This new parametrization allowed us to carry out successive observational analyses by decreasing its degrees of freedom systematically until ending up with a dynamical dark energy model that has the same number of parameters with $\Lambda$CDM. We found that the dark energy source with a dynamical equation of state parameter equal $-2/3$ at the early universe and $-1$ today fits the data slightly better than $\Lambda$. }
\\
\noindent
\hrule
\noindent \small{
\def\thefootnote{\arabic{footnote}}
\setcounter{footnote}{0}
\let\thefootnote\relax\footnote{\textbf{E-Mail:} oakarsu@ku.edu.tr, tdereli@ku.edu.tr, jvazquez@bnl.gov}
\def\thefootnote{\arabic{footnote}}
\setcounter{footnote}{0}}

\section{Introduction}
\label{sec:intro}
The recent high precision data is in very good agreement with the six parameter base $\Lambda$CDM 
cosmology \cite{WMAP11a,Planck13}. However, we still cannot consider it as the final cosmological model. 
It may represent, rather, a very good approximation or a limiting case of a more general theory, 
which does not necessarily reduce to vacuum energy in terms of a cosmological constant $\Lambda$ 
and may deviate from $\Lambda$ considerably in the far future and/or past. 
Two well known problems related with the $\Lambda$ assumption, the so called coincidence and fine-tuning problems, 
may be signs for a dynamical nature of dark energy \cite{Zeldovich,Weinberg89,Sahni00,Peebles03,Copeland06,Bamba12}.
Some tensions between the $\Lambda$ assumption and high precision data could be resolved in case of 
evolving dark energy (DE) (see for instance \cite{Sahni14,BOSS14,Vazquez12} and references therein). 

Using the one parameter extension to the $\Lambda$CDM, $w$CDM adopts a spatially flat universe and constant equation of state (EoS) parameter $w$ for DE, Planck collaboration \cite{Planck13} gives $w=-1.13_{-0.25}^{+0.23}$ from combined Planck+WP+highL+BAO data and BOSS collaboration \cite{BOSS14} gives $w=-0.97\pm0.05$ from the most recent combined Planck+BAO+CMB data. 
Constraining the possible evolution of $w$ on the other hand is difficult even with the powerful BAO, SN and CMB data that range at 
different redshift values. Hence, probing the possible DE evolution usually involves the introduction of a
 phenomenological parametrization for its EoS parameter involving a couple of free parameters. 
 The most widely used parametrization for constraining evolving DE is the linear expansion in scale factor $a$
 which is known as  CPL (Chevallier-Polarski-Linder \cite{CPL01,Linder03}) parametrization: 
 $w(a)=w_0+w_a (1-a)$, where $w_0$ and $w_{a}$ are real constants. It has been first 
 proposed as an alternative to the linear redshift parametrization of DE: $w(z)=w_{0}+w_{z} z$, 
 which in contrast to CPL grows increasingly unsuitable at redshifts $z\gg1$ and hence cannot be used with 
 high-redshift data, e.g., CMB. However, even the tightest constraints on the parameters of CPL from combined Planck+BAO+SN allow a quite flexible range for the evolution of the EoS parameter for DE \cite{BOSS14}. Besides this, compared to $\Lambda$CDM, the improvement in the success of fit to the data using this parametrization is not significant \cite{Planck13,BOSS14,Vazquez12}. These might be signaling for that CPL is not an adequate choice for describing the possible evolution of dark energy. It is already mentioned in many studies that the divergent behavior of CPL not only prevents one to make plausible predictions on the future of the universe but also shows that it cannot genuinely cover theoretical models of dark energy. 
 Throughout the literature there have been various DE parametrizations introduced by considering different reasoning and strategies for obtaining mathematically well behaving and physically acceptable models to obtain tighter constraints on DE and make plausible predictions on the future of the universe \cite{Bassett02,Jassal04,Gong07,FNT,Hannestad,Wetterich,Barboza08,Ma11,Sendra11,Zhang11,Bassett04,Shafieloo09,Wei13,Hazra15}.

In this paper we shall use a new parametrization for describing the DE source that may be taken as a 
natural extension of the same idea that give rise to linear redshift and  CPL parametrizations that correspond to the first two terms of Taylor expansion in $z$ and $a$, respectively. 
We start by introducing a fluid with an EoS parameter linear in time $\tilde{t}$. Then, we write its EoS parameter in terms of the scale factor $\tilde{a}$ by considering general relativity and utilize the obtained $w(\tilde{a})$ for describing the DE component of the universe in our observational analyses. We carry out successive observational analyses by decreasing the number of degrees of freedom (DoF) until we end up 
 with a dynamical DE model having no additional parameters compared to $\Lambda$CDM.

\section{Dark energy parametrization}
In this section we first derive an EoS parameter in terms of scale factor $\tilde{a}$, which yields linear EoS parameter in time $\tilde{t}$ when the universe is filled only with this fluid (here tilde denotes a universe contains only the fluid described by the EoS linear in time $\tilde{t}$). We then in Sec. \ref{sec:pde} make use of this form of $w(\tilde{a})$ for describing the DE source in the physical universe that contains sources other than DE also, by adopting $\tilde{a}\rightarrow a$, where $a$ represents the scale factor of the physical universe.

\subsection{Equation of state parameter linear in time}
Let us consider a fluid described by an EoS parameter expressed as a first order Taylor expansion in time $\tilde{t}$, that is:
\begin{equation}
\label{eqn:lteos}
w=w_{0}+w_{1}(1-\tilde{t}),
\end{equation}
where $w_0$ and $w_1$ are real constants and $\tilde{t}>0$ is the normalized time. It should be noted here that $w$ would not diverge provided that the time is restricted as in the big rip\footnote{See \cite{Caldwell03} for big rip cosmology.} cosmologies. We construct a relation between the fluid described by the EoS parameter linear in time \eqref{eqn:lteos}  and the scale factor by considering the general theory of relativity. Accordingly, we consider the Friedmann equations for spatially flat RW spacetime in the presence of a single fluid written as follows: 
$3\frac{\dot{\tilde{a}}^2}{\tilde{a}^2}=\kappa\rho$ and $\frac{\dot{\tilde{a}}^2}{\tilde{a}^2}+2\frac{\ddot{\tilde{a}}}{\tilde{a}}=-\kappa p=-\kappa \rho w$, 
where a dot denotes derivative with respect to time $\tilde{t}$. Eliminating $\rho$ between these two equations and then using \eqref{eqn:lteos} we get the corresponding
deceleration parameter\footnote{It may be noteworthy that a deceleration parameter linear in time for the observed universe was obtained from a higher dimensional cosmological model in dilaton gravity in \cite{AkarsuDilaton}, where its value at $\tilde{t}=0$ is depending on the number of extra dimensions.}
\begin{equation}
\tilde{q}\equiv -\frac{\ddot{\tilde{a}}\tilde{a}}{\dot{\tilde{a}}^2}=\frac{1}{2}+\frac{3}{2}\left[w_{0}+w_{1}(1-\tilde{t})\right],
\end{equation}
whose solution, for $w_1\neq0$, reads
\begin{equation}
\label{eqn:sfgeneral}
\tilde{a}=\tilde{a}_{1} \exp \left[ \frac{4}{3}\frac{{\rm arctanh}\left(       \frac{w_1 \tilde{t}-(w_0+w_1+1)}{\sqrt{(w_0+w_1+1)^2-c_1 w_1}}           \right)}{\sqrt{(w_0+w_1+1)^2-c_1 w_1}}  \right],
\end{equation}
where $\tilde{a}_{1}$ and $c_1$ are integration constants. For the case $w_1=0$, we would get the standard power law solution as $\tilde{a}\propto \tilde{t}^{2/3(1+w_0)}$ with $\rho\propto \tilde{a}^{-3(1+w_0)}$.
We demand a big bang $\tilde{a}=0$ at $\tilde{t}=0$, and expanding universe $\dot{\tilde{a}}\geq0$ for $\tilde{t}\geq0$ 
and $\tilde{a}\rightarrow+\infty$ at a finite time, say, as $\tilde{t}\rightarrow \tilde{t}_{\rm BR}$ where $\tilde{t}_{\rm BR}$ 
is the big rip time, so the EoS parameter never diverges.
One may check that the first condition is satisfied by choosing $c_1=0$, and in addition to 
this the latter two conditions imply the following relations $2+2w_0+2w_1>w_{1}>0$. 
Accordingly, substituting $c_1=0$ in \eqref{eqn:sfgeneral}, our solution reduces to
\begin{equation}
\label{eqn:sf}
\tilde{a}=\tilde{a}_{1} \exp \left[ \frac{4}{3}\frac{{\rm arctanh}
\left(       \frac{w_1 \tilde{t}}{w_0+w_1+1}-1           \right)}{w_0+w_1+1}                  \right],
\end{equation}
with $2+2w_0+2w_1>w_{1}>0$. Isolating $\tilde{t}$ in this solution by setting $\tilde{a}(\tilde{t}=1)=1$ we obtain
\begin{equation}
\label{eqn:lineartina}
\tilde{t}(\tilde{a})=\frac{2(1+w_0+w_1)}{w_1+(2+2w_0+w_1) \tilde{a}^{\frac{-3(1+w_0+w_1)}{2}}}.
\end{equation}
Finally using $\tilde{t}(\tilde{a})$ in \eqref{eqn:lteos} we obtain the following EoS in scale factor $\tilde{a}$
\begin{equation}
\label{eqn:Lint}
w(\tilde{a})=w_0+w_1\left[1-\frac{2(1+w_0+w_1)}{w_1+(2+2w_0+w_1) \tilde{a}^{\frac{-3(1+w_0+w_1)}{2}}}\right]
\end{equation}
with $2+2w_0+2w_1>w_1>0$. Next, using the energy-momentum conservation equation 
we obtain the energy density $\rho$:
\begin{equation}
\label{rhoproper}
\rho=\rho_{0} \tilde{a}^{-3(w_{0}+w_{1}+1)} \left(1+w_1\frac{\tilde{a}^{\frac{3}{2}(w_{0}+w_{1}+1)}-1}{2w_{0}+2w_{1}+2}\right)^4,
\end{equation}
where $\rho_{0}=\rho(\tilde{a}=1)$. We notice that 
$w\rightarrow \frac{-3(w_0+1)\ln(\tilde{a})-4w_0}{3(w_0+1)\ln(\tilde{a})-4}$ and 
$\rho\rightarrow \rho_{0} \left[-1+\frac{3(w_0+1)}{4} \ln (\tilde{a}) \right]^4$ as $w_{1}\rightarrow-1-w_{0}$.

The EoS parameter \eqref{eqn:Lint} 
is bounded both as $\tilde{a}\rightarrow0$ and $\tilde{a}\rightarrow\infty$, namely, $w\rightarrow w_{0}+w_1$ 
as  $\tilde{a}\rightarrow 0$ and $w\rightarrow -w_{0}-w_1-1$ as $\tilde{a}\rightarrow \infty$. 
We note that $\tilde{a}\rightarrow0$ as $\tilde{t}\rightarrow0$ and $\tilde{a}\rightarrow\infty$ as $\tilde{t}\rightarrow 2(1+w_0+w_1)/w_1$, hence the time could not take arbitrarily large values unless $w_1=0$,
which also explains why EoS parameter linear in $\tilde{t}$ does not diverge in future in this case.

The recent high precision data is in very good agreement with $\Lambda$ as the DE source in the universe. However, it may represent, rather, a very good approximation or a limiting case of a more general theory, which does not reduce necessarily to vacuum energy in terms of a cosmological constant. In this regard, being  $w_{\rm de}\sim -1$ is in very good agreement with precision data, it seems physically more acceptable to consider an approximate description of possibly dynamical DE with a parametrization that is  mathematically well behaved and may never depart from $w_{\rm de}\sim -1$ a lot. We note that the EoS parameter we derived has these properties and hence may be adopted for parametrizing the EoS of the DE source in the universe containing other sources also.

\subsection{Parametrization of dark energy source}
\label{sec:pde}

The EoS derived above was first given in \cite{Akarsu11} starting from the linear deceleration parameter in time using general theory of relativity and then utilized in \cite{Akarsu14,Kumar14} for describing the effective EoS parameter averaging all the ingredients of the universe to constrain particularly the kinematical properties of the expansion of the universe. In this work, on the other hand, we follow a quite different method and adopt the EoS written in terms of $\tilde{a}$ for describing
only the DE component of the universe, rather than the average ingredient of the universe. 
The cost of this is that \eqref{eqn:Lint} will not correspond to the linear EoS in physical time $t$ in the actual universe anymore, unless the universe is dominated by this fluid. The reason being is that including the other sources 
such as cold dark matter (CDM) in addition to DE the evolution of the scale factor will deviate from 
the one given in \eqref{eqn:sf}, i.e., the relation between the scale factor $a$ and time $t$ will be different, until the DE source becomes dominant over all other sources in the universe.

Let us now adopt the EoS parameter given in \eqref{eqn:Lint} for describing DE source by setting $\tilde{a}\rightarrow a$ and then write it in a more useful form by defining $w_{\rm i}=w_0+w_1$:
\begin{equation}
\label{eqn:eos2}
w_{\rm de}=w_{\rm i}-\frac{2(w_{\rm i}-w_0) (1+w_{\rm i})}{(w_{\rm i}-w_0)+(2+w_{\rm i}+w_0) \,a^{-\frac{3}{2}(1+w_{\rm i})}},
\end{equation}
which also recasts the energy density \eqref{rhoproper} as
\begin{equation}
\label{eqn:definal}
\rho_{\rm de}=\rho_{\rm de}^{(0)} \left[\frac{w_{\rm i}-w_0}{2w_{\rm i}+2} a^{\frac{3}{4}(w_{\rm i}+1)} + \frac{w_{\rm i}+w_0+2}{2w_{\rm i}+2} a^{-\frac{3}{4}(w_{\rm i}+1)}       \right]^{4},
\end{equation}
where $a$ is the physical scale factor, {{}}{and the conditions on the parameters are now given by $-2-w_{\rm i}<w_0<w_{\rm i}$}.  Our parametrization has the same number of degrees of freedom with the CPL but while the variation of the EoS of DE with respect the scale factor is 
constant in scale factor $w_{\rm CPL}'=-w_{a}$ in CPL, here it is a dynamical quantity:
\begin{equation}
w_{\rm de}'=\frac{3(w_{\rm i}+1)^2[ (w_0+1)^2-(w_{\rm i}+1)^2 ] a^{\frac{3}{2}w_{\rm i}+\frac{1}{2}}}{\left[(w_0-w_{\rm i})a^{\frac{3}{2}w_{\rm i}+\frac{3}{2}}-(w_0+w_{\rm i}+2)\right]^2},
\label{eqn:wprime}
\end{equation}
where a prime denotes derivative with respect to the scale factor $a$.

{{}}{We note that the form of the energy density \eqref{eqn:definal} we obtained by the reasoning we followed in the previous section  has a symmetry for $w_{\rm i}+1 \to -w_{\rm i}-1$ and hence it is enough if we consider the cases $w_{\rm i}>-1$ only. In accordance with this, we see from \eqref{eqn:eos2} that $w_{\rm de}\rightarrow  w_{\rm i}$ as $a\rightarrow 0$ ($z\rightarrow \infty$) and hence $w_{\rm i}$ denotes the initial value of the EoS parameter of DE. Similarly $w_{\rm de}\rightarrow  -w_{\rm i}-2$ as 
$a\rightarrow\infty$ ($z\rightarrow -1$), and hence we can write the final value of the EoS parameter as $w_{\rm f}= -w_{\rm i}-2$. We also note that $w_{\rm de}'$ can take only negative values as long as we stick to the range $-w_{\rm i}-2<w_0<w_{\rm i}$ for the parameters, but take positive values if this condition could be violated. The radiation and pressure-less matter components of the universe enforce universe to expand as $a\propto t^{\frac{1}{2}}$ and $a\propto t^{\frac{2}{3}}$, respectively. The DE source described by \eqref{eqn:definal}, on the other hand, enforces universe to expand according to the kinematics we discussed in the previous section. In the physical universe, say, in the presence of all these components, the expansion of the universe will be much complicated as it will be determined by the joint effect of these components. We also note that the conditions on the parameters of the hypothetical fluid, we are now utilizing for describing DE, were initially considered to obtain a physically reasonable expansion history for the universe filled by this fluid only. However the presence of radiation and pressure-less matter in addition to a DE source described by \eqref{eqn:definal} can lead to viable expansion histories for the universe even if the parameters $w_{\rm i}$ and $w_{0}$ take values out of the range $-w_{\rm i}-2<w_0<w_{\rm i}$ and hence provides us freedom to violate this condition. Accordingly, in the next section we will carry out observational analyses without taking this condition into account, so that our analyses will not be restricted to the cases $w_{\rm de}'<0$. However we will see below that the concluding values for $w_{\rm i}$ and $w_{0}$ upon the three successive observational analyses, where we decrease the number of degrees of freedom systematically, satisfy the condition $-w_{\rm i}-2<w_0<w_{\rm i}$.}


\section{Observational constraints}
\label{sec:obs}

We consider a spatially flat RW spacetime and the Hubble parameter $H$ in the presence of radiation $\Omega_{\rm r}$, 
pressure-less fluid/(CDM+matter) $\Omega_{\rm m}$  and the dark energy $\Omega_{\rm de}$ described by \eqref{eqn:definal} is given by
\begin{align}
\frac{H^2}{H_0^2}=\Omega_{\rm r}^{(0)} (1+z)^{4}+\Omega_{\rm m}^{(0)} (1+z)^{3}+\Omega_{\rm de}^{(0)}  \left[\frac{w_{\rm i}-w_0}{2w_{\rm i}+2} (1+z)^{-\frac{3}{4}(w_{\rm i}+1)} + 
\frac{w_{\rm i}+w_0+2}{2w_{\rm i}+2} (1+z)^{\frac{3}{4}(w_{\rm i}+1)}       \right]^{4},
\end{align}
where the density parameter of the DE $\Omega_{\rm de}^{(0)}=1-\Omega_{\rm m}^{(0)}-\Omega_{\rm r}^{(0)}$ {{}}{and $w_{\rm i}>-1$}. Here subscript/superscript $0$ indicates the values of the parameters today. The density parameter of radiation is 
$\Omega_{\rm r}^{(0)}=2.469\times 10^{-5} h^{-2}(1+0.2271 N_{\rm eff})$, where 
$h=H_0/100 {\rm km s}^{-1}{\rm Mpc}^{-1}$ \cite{WMAP11a}. We consider a model with standard matter and radiation content, including three neutrino species with minimum allowed mass 
$\sum m_{\nu}=0.06\, {\rm eV}$. 
 {{}}{Throughout the analysis we assume flat priors over our sampling parameters: $w_{0}=[-2.0,-0.5]$, $w_{\rm i}=[-1.0,2.0]$ for the dark energy EoS parameters, $\Omega_{\rm m}^{(0)}=[0.05,1]$ for the pressure-less matter density parameter today, $\Omega_{\rm b}^{(0)} h^2=[0.02,0.025]$ for the baryon density today and $h=[0.4,1.0]$ for the reduced Hubble constant.} {{}}{It might be noteworthy to note that the radiation energy density $\rho_{\rm r}^{(0)}=\Omega_{\rm r}^{(0)} H_0^2$ today is not subject to our analysis since it is well constrained, such that it has a simple $\rho_{\rm r}=\frac{\pi^2}{15} T_{\rm CMB}^4$ relation with the CMB monopole temperature (see \cite{Dodelson03} for further details), which today is very precisely measured to be $T_{\rm CMB}^{(0)}=2.7255\pm 0.0006\,{\rm K}$ \cite{Fixsen09}.}
 
In order to perform the parameter space exploration, we make use of a modified version of 
a simple and fast Markov Chain Monte Carlo (MCMC) code that computes expansion
rates and distances from the Friedmann equation, 
named SimpleMC.
This code already contains a compressed version of the Planck data, a recent reanalysis of Type Ia supernova (SN) data,
and high-precision BAO measurements at different redshifts up to $z<2.36$ \cite{BOSS14}.

In three successive subsections we carry out observational analyses:
i) $w_0$ and $w_{\rm i}$ are both free parameters: DoF are the same with CPL, ii) $w_0$ is free but
 $w_{\rm i}$ is fixed to certain values, so that DoF are the same with $w$CDM,
  and iii) both $w_0$ and $w_{\rm i}$ are fixed to certain values so that  DoF are the same with $\Lambda$CDM.

\subsection{General case: free $w_0$ and $w_{\rm i}$}

\begin{figure}[h!]\centering
\includegraphics[width=0.27\textwidth]{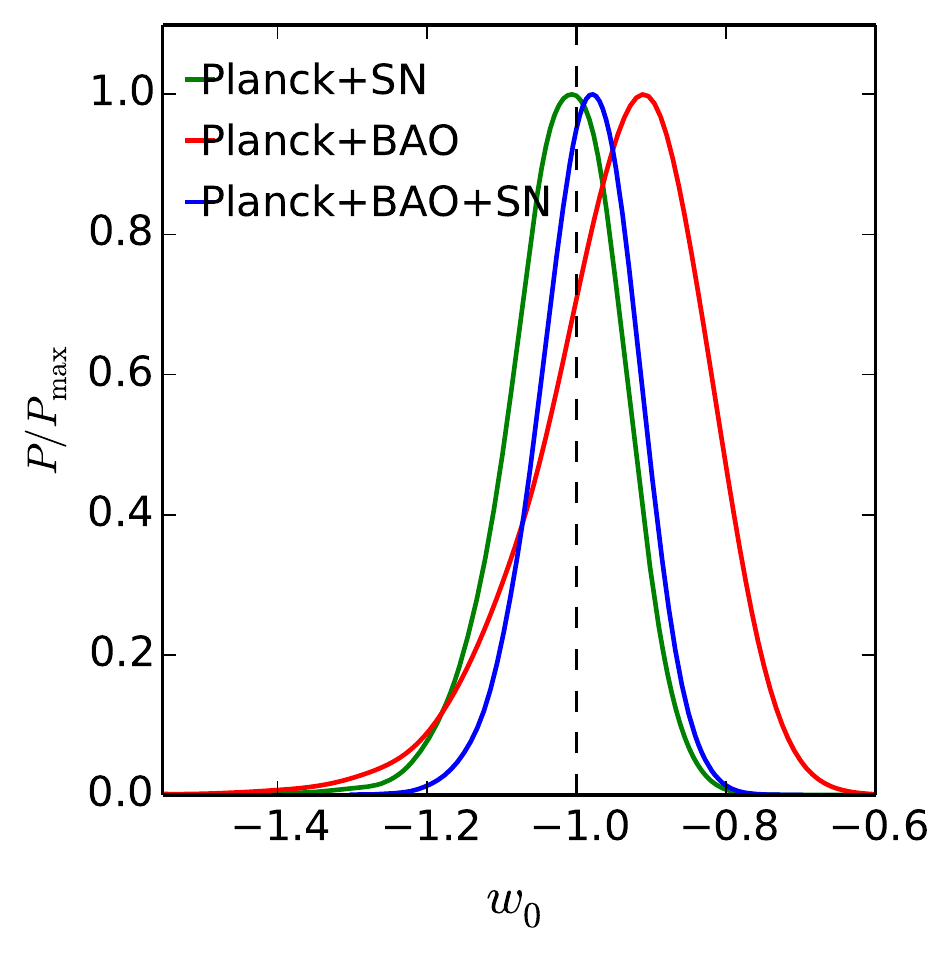}
\includegraphics[width=0.27\textwidth]{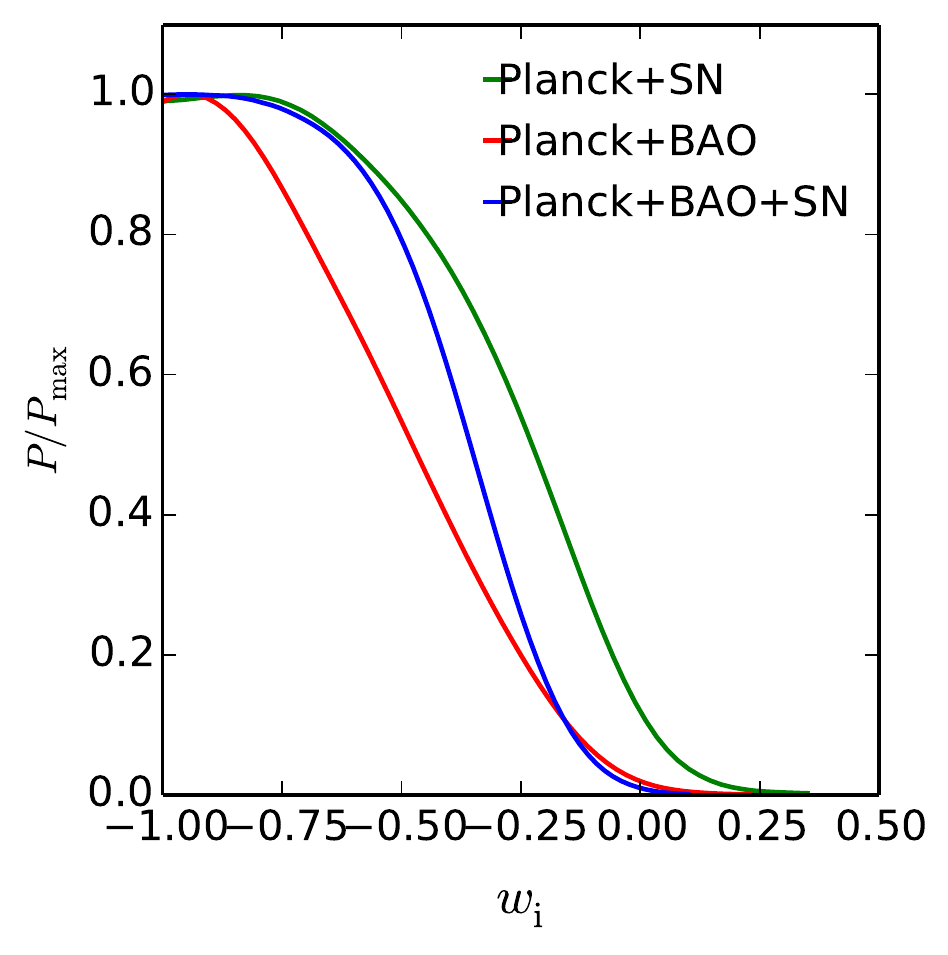}
\includegraphics[width=0.37\textwidth]{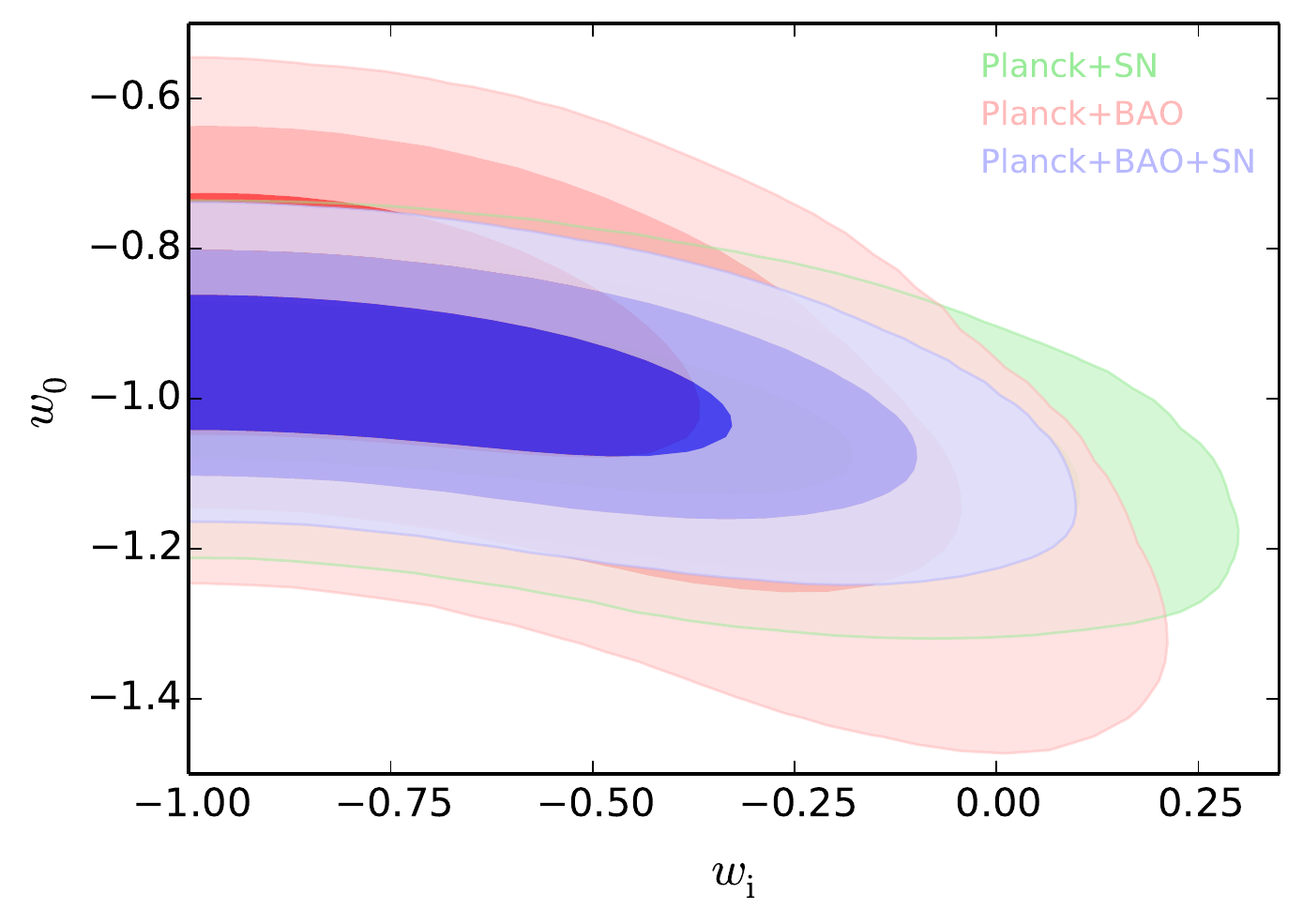}
\caption{1-D and 2-D posterior distributions for the parameters $w_0$ and $w_{\rm i}$;  data combinations are indicated in the legends.}
\label{fig:probw0wi}
\end{figure}

\begin{figure}[h!]\centering
\includegraphics[width=0.40\textwidth]{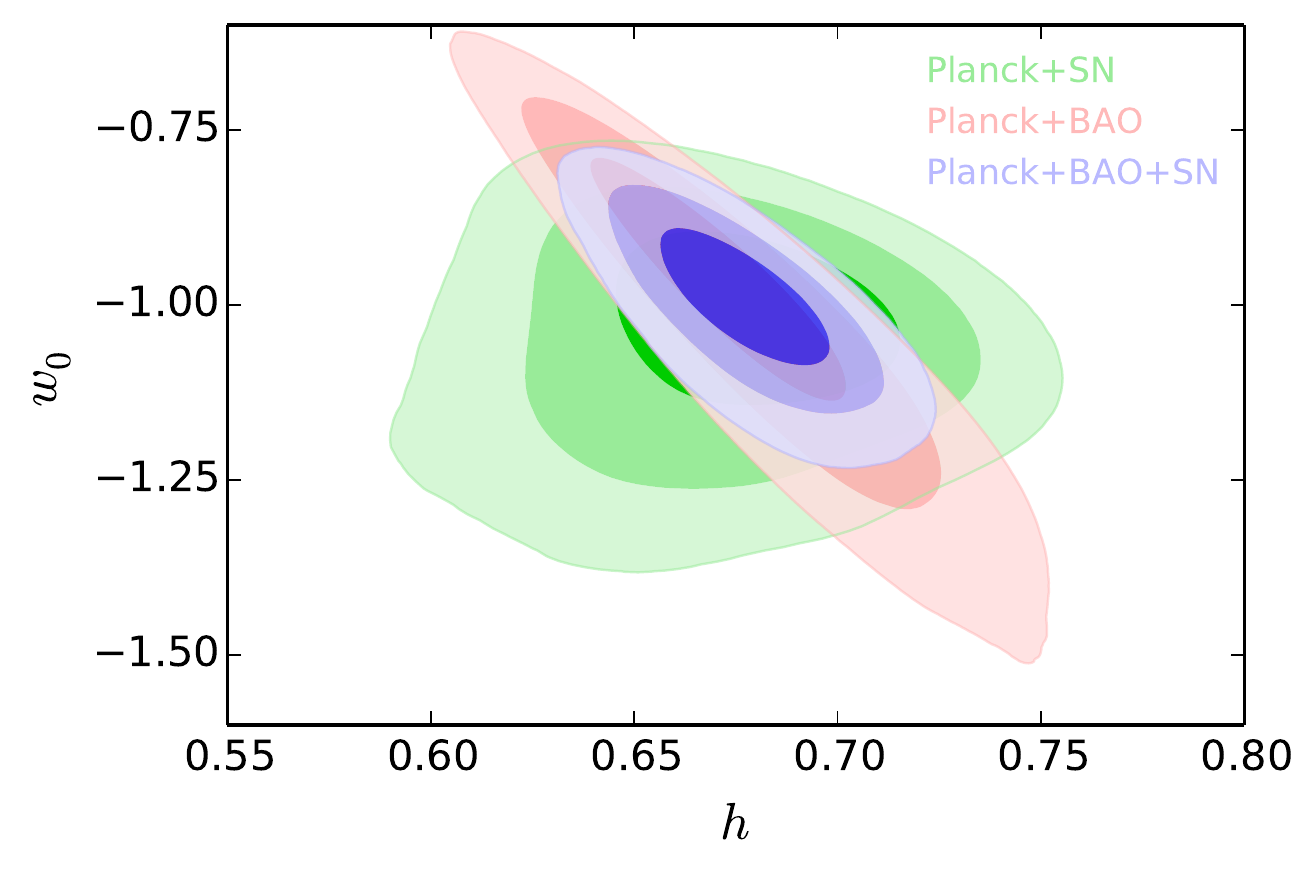}
\includegraphics[width=0.40\textwidth]{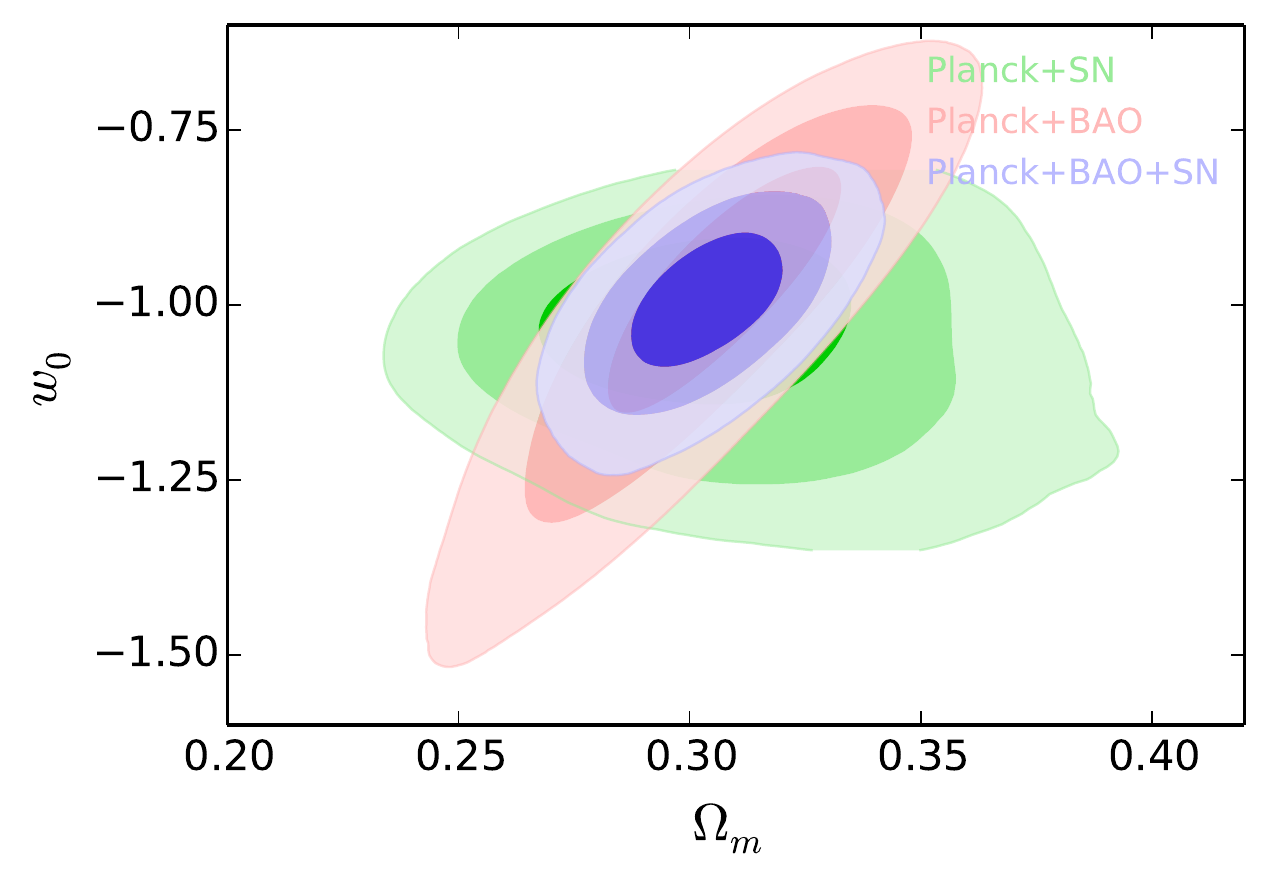}
\caption{2-D posterior distributions for $w_0-h$ and $w_0-\Omega_{\rm m}^{(0)}$; data combinations are indicated in the legends.}
\label{fig:w0HandOm}
\end{figure}

We first constrain the cosmological parameters by taking both $w_0$ and $w_{\rm i}$ 
as free parameters so that we have the same number of DoF with the CPL parametrization. We summarize the results for the data sets Planck+BAO, Planck+SN and Planck+BAO+SN in 
Table \ref{table:LVDPobs}.
In Figure \ref{fig:probw0wi} we show 1-D {{}}{and 2-D} posterior distributions of $w_{0}$ and $w_{\rm i}$ 
for the following data sets Planck+BAO, Planck+SN and Planck+BAO+SN. In Figure \ref{fig:w0HandOm} we give 
2-D posterior distributions for $w_0-h$ and $w_0-\Omega_{\rm m}^{(0)}$.


\begin{table}[h!]\centering\small
\caption{Parameter constraints. SN data cannot constraint 
$w_{\rm i}$ as good as $w_{\rm 0}$ while it is the other way around for BAO data. For two-tailed distributions the results are given in $1\sigma$ and for one-tailed distributions given in $2\sigma$.} 
\begin{tabular}{lcccc}
\hline\hline Data set & $\Omega^{(0)}_{{\rm m}}$ & $h$ & $w_{0}$ & $-1<w_{\rm i}$   \\ \hline\hline\\
Planck+BAO & $0.306^{+0.014}_{-0.015}$ &  $0.673^{+0.020}_{-0.018}$ & $-0.97\pm 0.10$ &  $< -0.30$ \\[8pt]
Planck+SN &  $0.304^{+0.021}_{-0.019}$ & $0.677\pm 0.019$ &   $ -1.02^{+0.07}_{-0.08}$ & $< -0.27$ \\[8pt]
\textbf{Planck+BAO+SN} & $0.304\pm 0.009$ & $0.677 \pm 0.011$ & $ -0.99\pm 0.06$ & $< -0.42$ \\[8pt]
\hline\hline
\end{tabular}
\label{table:LVDPobs}
\end{table}

We observe that the matter density parameter $\Omega_{\rm m}^{(0)}$ and the reduced Hubble constant $h$ 
are obtained almost the same in three combinations of data sets and Planck+BAO+SN data set leads to 
the tightest constraints; $\Omega_{\rm m}^{(0)}=0.304\pm 0.009$ and $h=0.677 \pm 0.011$.

In all cases the best-fit parameters are consistent with $\Lambda$CDM model, i.e., 
$(w_0,w_{\rm i})=(-1,-1)$. We observe that Planck+SN data lead to a tighter constraint on $w_0$ 
while Planck+BAO data lead a tighter constraint on $w_{\rm i}$. 
The central value of $w_0$ is almost equal to $-1$ from both Planck+BAO, Planck+SN and Planck+BAO+SN and its 
value from the different data combinations changes by less than $1\sigma$. Using three data sets at the 
same time we obtain $w_0= -0.99\pm 0.06$ (Planck+BAO+SN) with $1\sigma$ having the tightest constraint on $w_0$ 
with a central value almost exactly equal to $-1$. Using CPL, on the other hand, $w_0$ is not 
constrained very well and the central value of $w_0$ is obtained not only quite higher than $-1$ but also 
significantly different from three different data combinations; $w_0=-0.58\pm0.24$ (Planck+BAO), 
$w_0=-0.90\pm0.16$ (Planck+SN) and $w_0=-0.93\pm0.11$ (Planck+BAO+SN) \cite{BOSS14}. 
Constraint on $w_{\rm i}$ from Planck+SN is quite loose such that $-1<w_{\rm i}<-0.27$ ($2\sigma$); so that the DE could either behave like cosmic strings or a cosmological constant at 
very high red-shift values. Planck+BAO data constrain $w_{\rm i}$ better and constrict the allowed 
range to $-1<w_{\rm i}<-0.30$ ($2\sigma$). The tightest constraint is obtained from Planck+BAO+SN data; $-1<w_{\rm i}<-0.42$ ($2\sigma$). According to this initially dark matter and cosmic strings like DE scenarios are not viable. Considering the constraints on $w_0$ and $w_{\rm i}$ together we observe 
that $\Lambda$ as the DE source is doing great but even considerably large deviations from $\Lambda$
are still allowed, for instance, $(w_0,w_{\rm i})=(-1,-\frac{2}{3})$ is also perfectly allowed in this picture. 
The best fitting model to Planck+BAO+SN data has $\chi^2=46.63$, representing 
an improvement of $\Delta\chi^2=0.37$ compared to $\Lambda$CDM for which 
$\chi^2_{\Lambda{\rm CDM}}=47.00$, for the addition of two extra parameters. 
This is an improvement almost the same with the one in the CPL model $\Delta\chi^2_{\rm CPL}=0.42$ \cite{BOSS14}. Accordingly in terms of information criteria the improvement is not sufficient enough to 
justify the addition of two extra degrees of freedom either in our parametrization  or CPL parametrization 
and additionally there is no reason to prefer one over the other among these two parametrizations.

\subsection{Free $w_0$ and fixed $w_{\rm i}$}

\begin{figure}[h!]\centering
\includegraphics[width=0.30\textwidth]{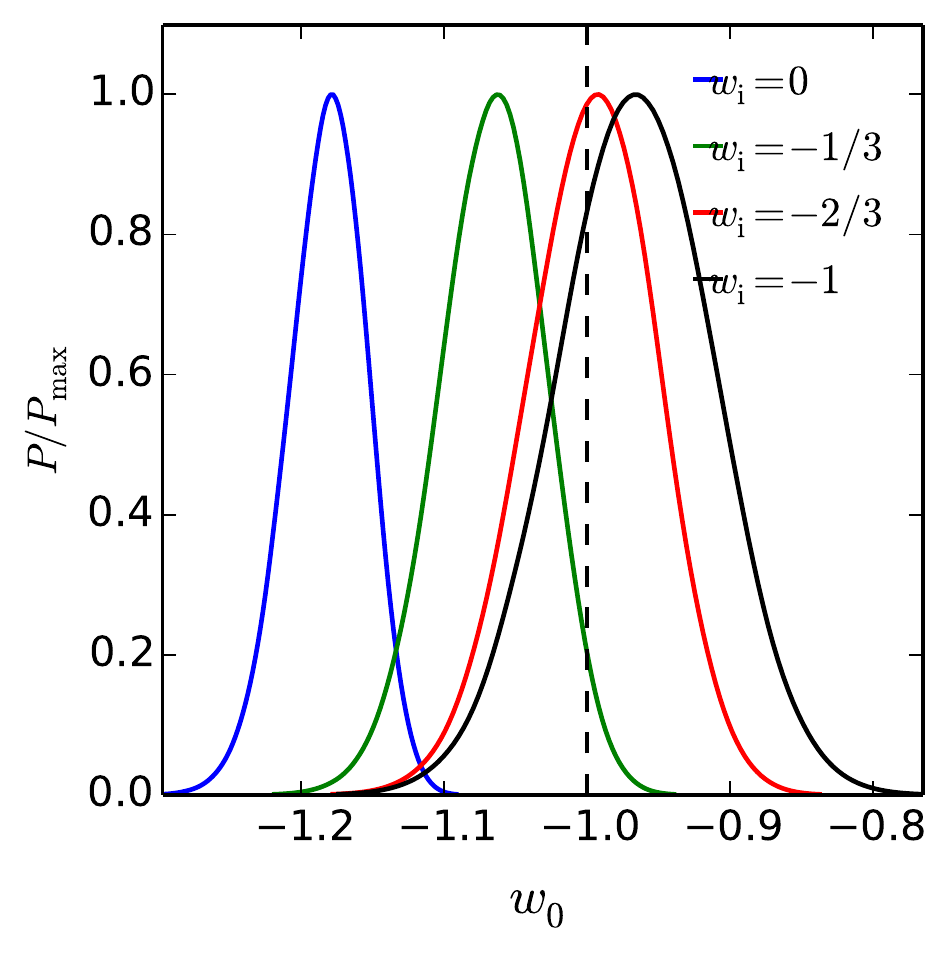}
\caption{1-D posterior distributions for $w_0$ for the particular values $w_{\rm i}=0$ (blue), $w_{\rm i}=-\frac{1}{3}$ (green), $w_{\rm i}=-\frac{2}{3}$ (red) and $w_{\rm i}=-1$ (black) from Planck+BAO+SN data.}
\label{fig:probredm}
\end{figure}

We may try to get more information using our parametrization by reducing its 
DoF to the that of the $w$CDM parametrization by fixing $w_{\rm i}$. We now carry out the observational analyses using 
Planck+BAO+SN data by setting either $w_{\rm i}=0$ (DE starts like pressure-less matter), $w_{\rm i}=-\frac{1}{3}$ 
(DE starts like cosmic strings) or $w_{\rm i}=-\frac{2}{3}$ (DE starts like cosmic domain walls). In Table 
\ref{table:LVDPobs2}, we summarize the results, including the minimum $\chi^2$ values, from Planck+BAO+SN data. 
For a comparison, the constraint on $w$CDM 
parametrization from Planck+BAO+SN data is $w_0=-0.97\pm0.08$ ($1\sigma$) \cite{BOSS14}. We give the 1-D 
probability distributions of $w_0$ in our model in Figure \ref{fig:probredm}. We note that the central value of $w_0$ shifts from 
the values less than $-1$ to the values higher than $-1$ and the $\chi^2$ improves as $w_{\rm i}$ goes from 
$0$ to $-1$. Given that the case $w_{\rm i}=0$ yields very large $\chi^2$ relative to other three, it can be rejected. The most interesting point in Table \ref{table:LVDPobs2} is that $w_0$ takes the closest value to $-1$ in the case 
$w_{\rm i}=-\frac{2}{3}$ rather than in the case $w_{\rm i}=-1$ and these two cases have almost the same $\chi^2$ 
values that are considerably lower than the cases $w_{\rm i}=0$ and $w_{\rm i}=-\frac{1}{3}$. 
This may be interpreted as a signal for a DE with a dynamical EoS parameter starting from values $\sim -\frac{2}{3}$ 
at earlier times and then approaching a cosmological constant as the universe expands, as noted also by \cite{Vazquez12b}.

\begin{table}[h!]\centering\small
\caption{Mean values with $1\sigma$ errors of the parameters of the model for the particular values 
$w_{\rm i}=0$, $w_{\rm i}=-\frac{1}{3}$, $w_{\rm i}=-\frac{2}{3}$ and $w_{\rm i}=-1$ from Planck+BAO+SN data.} 
\begin{tabular}{ccccc}
\hline\hline $\Omega^{(0)}_{{\rm m}}$ & $h$ & $w_{0}$ & $w_{\rm i}$ & $\chi^2$   \\ \hline\hline\\
$0.289\pm 0.008$ & $0.688\pm 0.010$ & $-1.180_{-0.025}^{+0.028}$ & $0$ & $54.04$ \\[8pt] 
$0.298\pm 0.008$ & $0.680\pm 0.010$ & $-1.065_{-0.035}^{+0.037}$ & $-\frac{1}{3}$  & $48.44$ \\[8pt]
$0.303\pm 0.009$ & $0.677_{-0.011}^{+0.009}$ & $-0.995_{-0.042}^{+0.046}$ & $-\frac{2}{3}$ & $46.90$ \\[8pt]
$ 0.305\pm 0.009 $ & $ 0.676_{-0.011}^{+0.010} $ & $ -0.966\pm 0.053$ & $-1$ & $ 46.64 $ \\[8pt]
 
\hline\hline
\end{tabular}
\label{table:LVDPobs2}
\end{table}

\subsection{Fixed $w_0$ and fixed $w_{\rm i}$}

\begin{figure}[h!]\centering
\includegraphics[width=0.49\textwidth]{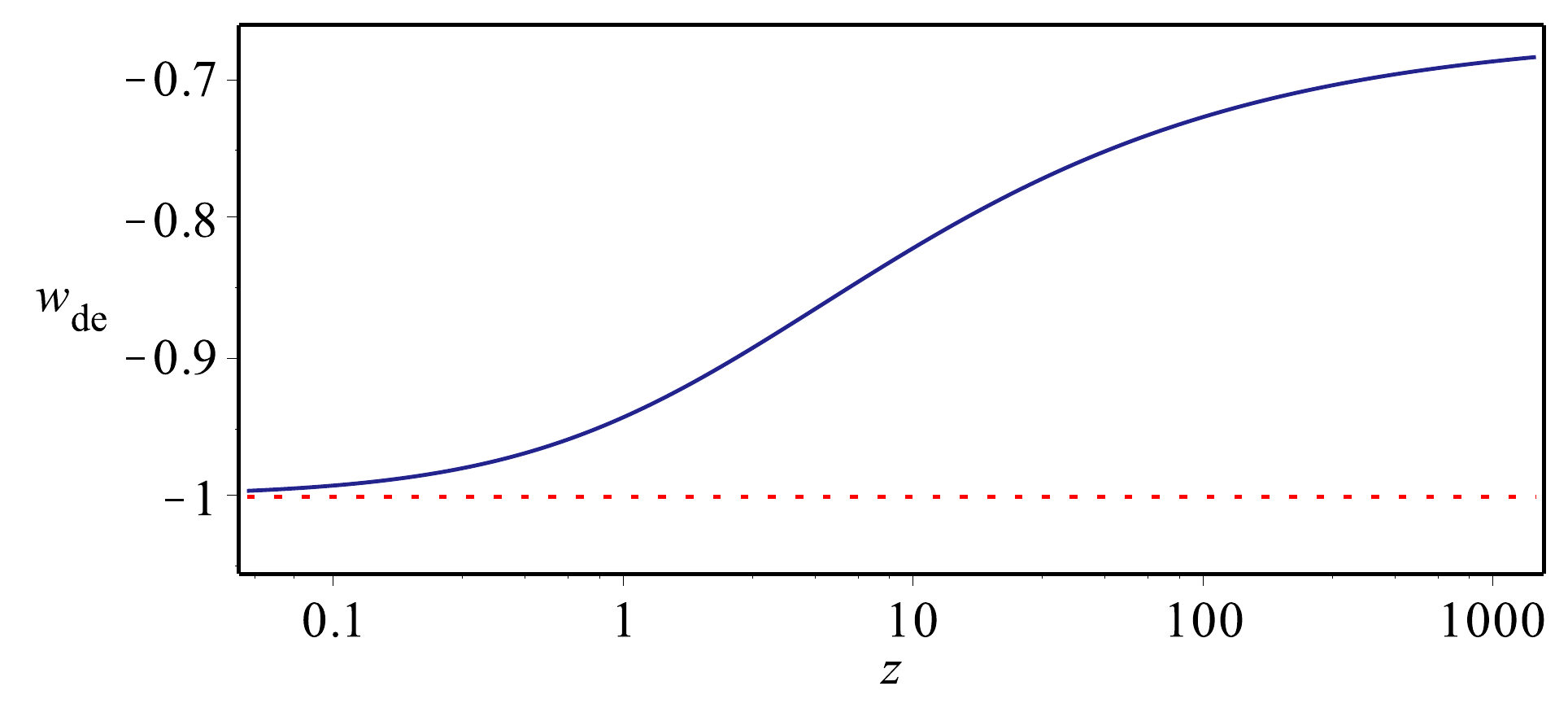}
\includegraphics[width=0.49\textwidth]{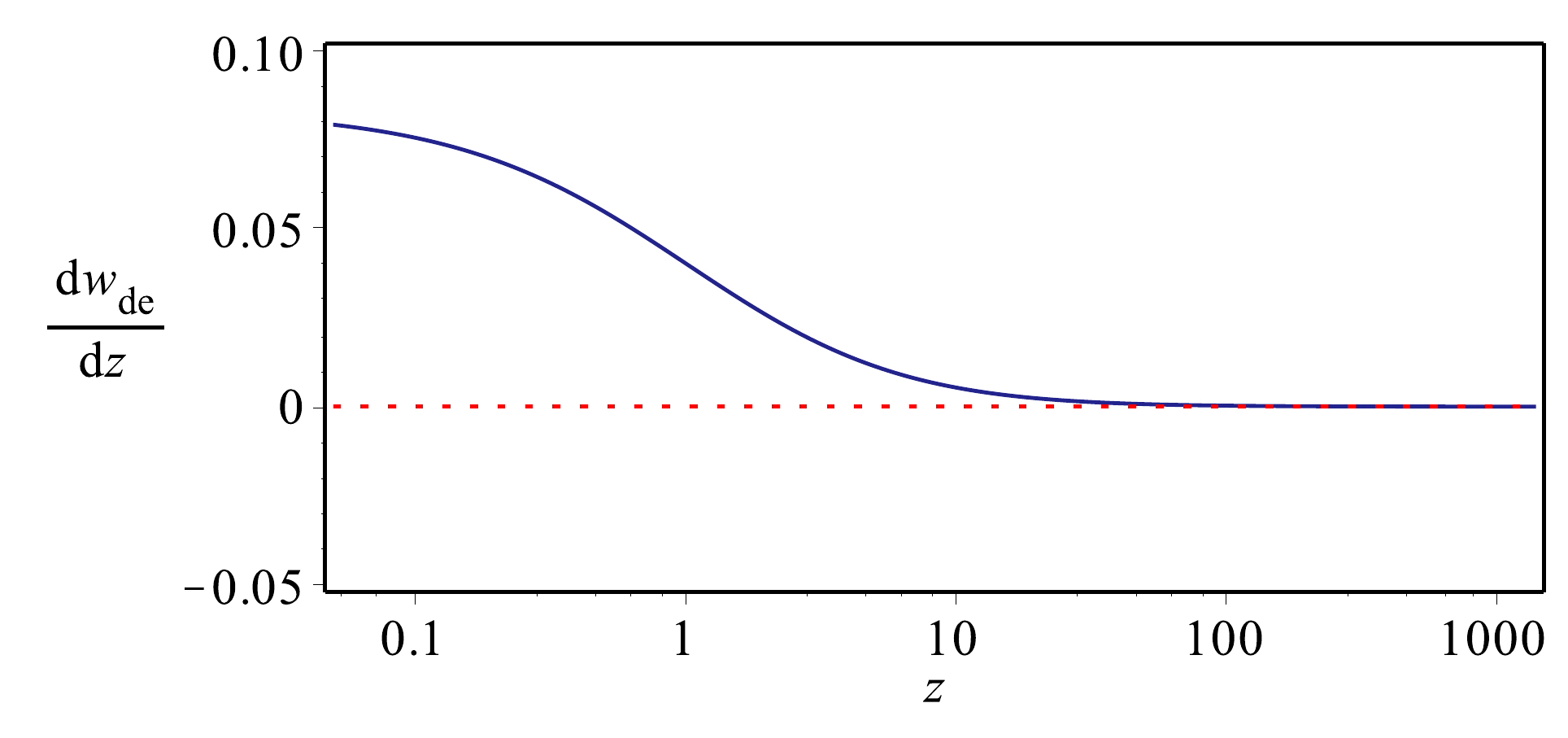}
\caption{$w_{\rm de}$ and ${\rm d} w_{\rm de}/{\rm d} z$ in terms of redshift $z$ for the case $(w_0,w_{\rm i})=(-1,-\frac{2}{3})$ (solid lines). The dotted lines correspond to the case cosmological constant.}
\label{fig:wevol}
\end{figure}

We observe from Table \ref{table:LVDPobs2} that $w_0$ persists on yielding values around $-1$ in spite of the large differences between the fixed values of $w_{\rm i}$. Hence we now fix also $w_{0}=-1$ along with $w_{\rm i}=0$, $w_{\rm i}=-\frac{1}{3}$, $w_{\rm i}=-\frac{2}{3}$ or $w_{\rm i}=-1$. Doing so our model now yields the same number of DoF with the $\Lambda$CDM model and the latter case $(w_0,w_{\rm i})=(-1,-1)$ corresponds to $\Lambda$ while the former three cases correspond to dynamical DE models. We summarize the constraints from Planck+BAO+SN data set with their minimum $\chi^2$ values in Table \ref{table:LVDPobs3}.
\begin{table}[h]\centering\small
\caption{Mean values with $1\sigma$ errors of the parameters of the model with  $w_{0}=-1$ for the particular values $w_{\rm i}=0$, $w_{\rm i}=-\frac{1}{3}$ and $w_{\rm i}=-\frac{2}{3}$ and $\chi^2$ values from Planck+BAO+SN data. The case $w_{\rm i}=-1$ corresponds to $\Lambda$CDM.} 
\begin{tabular}{ccccc}
\hline\hline $\Omega^{(0)}_{{\rm m}}$ & $h$ & $w_{0}$ & $w_{\rm i}$ & $\chi^2$   \\ \hline\hline\\
 $0.316_{-0.005}^{+0.006}$ & $0.605\pm 0.002$ & $-1$ & $0$ & $173.34$ \\[8pt]
 $0.305_{-0.008}^{+0.007}$ & $0.664\pm 0.005$ & $-1$ & $-\frac{1}{3}$  & $51.80$ \\[8pt]
 $0.302_{-0.008}^{+0.007}$ & $0.678\pm 0.006$ & $-1$ & $-\frac{2}{3}$ & $46.88$ \\[8pt]
 $0.302\pm 0.008$ & $0.682\pm 0.006$ & $-1$ & $-1$ & $47.00$ \\[8pt]
\hline\hline
\end{tabular}
\label{table:LVDPobs3}
\end{table}
We note that the case $(w_0,w_{\rm i})=(-1,0)$ and $(w_0,w_{\rm i})=(-1,-\frac{1}{3})$ 
should be ruled out due to their very large $\chi^2$ values. We note on the other hand that the cases 
$(w_0,w_{\rm i})=(-1,-\frac{2}{3})$ and $(w_0,w_{\rm i})=(-1,-1)$, i.e. $\Lambda$CDM,
have almost the same low $\chi^2$ values; $\chi^2_{(-1,-2/3)}=46.88$ and $\chi^2_{\Lambda{\rm CDM}}=47.00$, 
respectively. 
{{}}{However, it is noteworthy to note the case $(w_0,w_{\rm i})=(-1,-\frac{2}{3})$ has the lowest $\chi^2$ value
with a difference over the $\Lambda$CDM given by $\chi^2_{\Lambda{\rm CDM}}-\chi^2_{(-1,-2/3)}=0.12$ , although not
statistically significant.} 
In this case we 
find from \eqref{eqn:wprime} that $w_{\rm de}'<0$ through the history of universe and $w_{\rm de}'(z=0)=-\frac{1}{12}$ 
today, while $w'$ is always null in the case of $\Lambda$. According to this we have a dynamical DE model that fits 
to data equally better than $\Lambda$ but yet does never cover/mimic $\Lambda$. We plot $w_{\rm de}$ and ${\rm d} w_{\rm de}/{\rm d} z=-w_{\rm de}'/(1+z)^2$ versus redshift $z$ in Fig. \ref{fig:wevol} for demonstrating how the dynamics of DE in case $(w_0,w_{\rm i})=(-1,-\frac{2}{3})$ deviate from $\Lambda$.

\begin{figure}[h!]\centering
\includegraphics[width=0.49\textwidth]{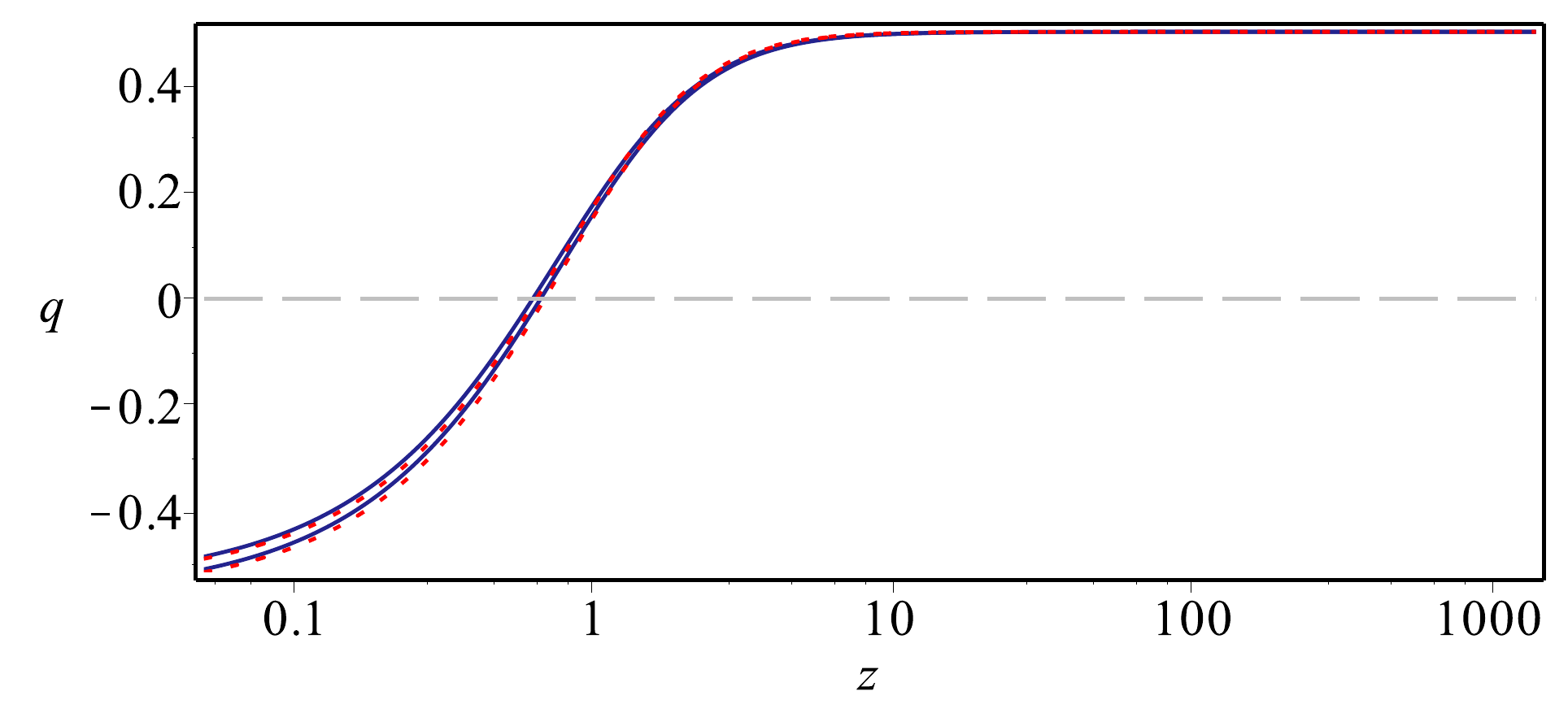}
\includegraphics[width=0.49\textwidth]{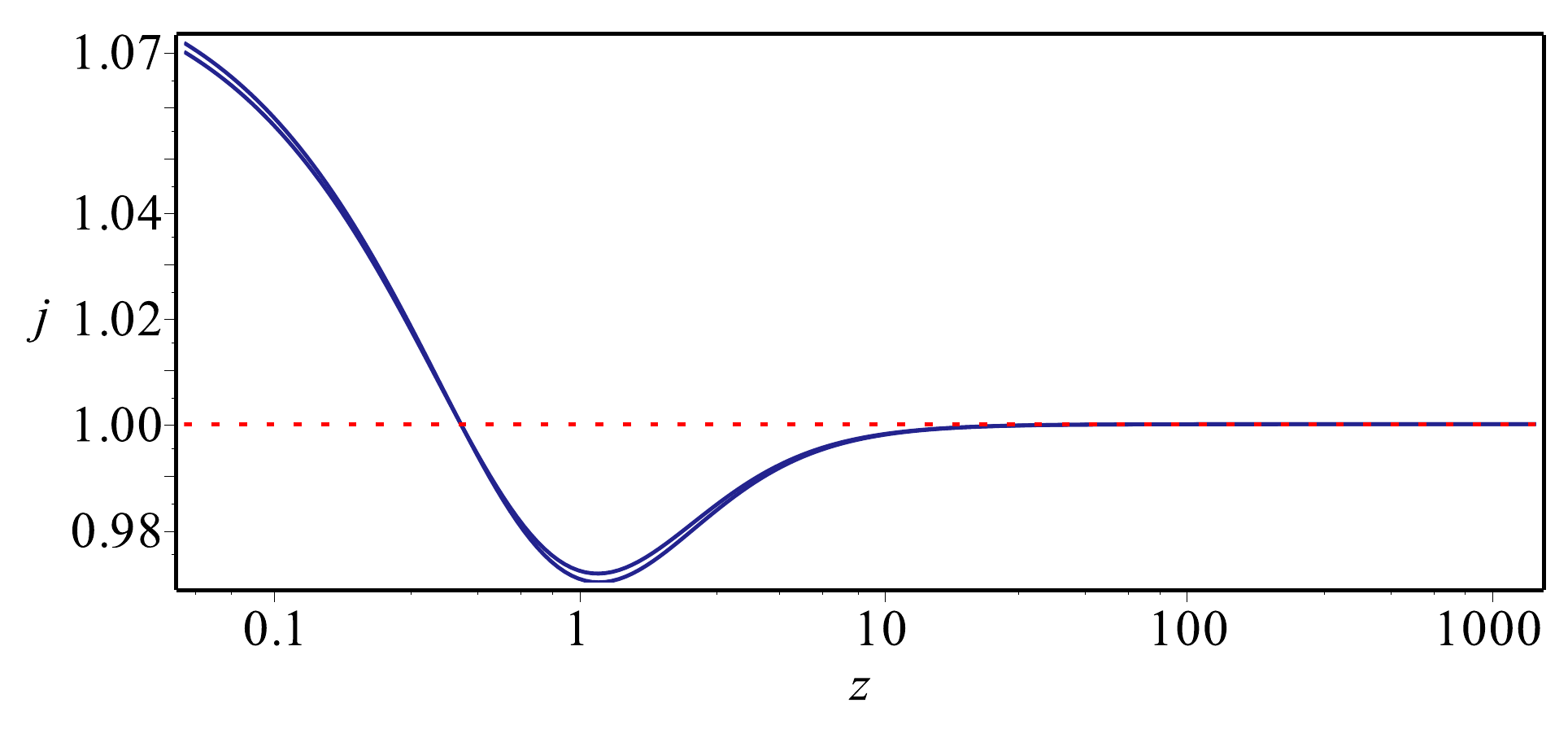}
\caption{Deceleration parameter $q$ and jerk parameter $j$ versus redshift $z$ for the case $(w_0,w_{\rm i})=(-1,-\frac{2}{3})$ (solid lines) and $\Lambda$CDM (dotted lines).}
\label{fig:qj}
\end{figure}

Two key parameters in cosmology are the deceleration and jerk parameters that are defined as $q=-\frac{\ddot{a}}{aH^2}=\frac{(1+z){\rm d}H}{H{\rm d}z}-1$ and $j=\frac{\dddot{a}}{aH^3}=q+2q^2+(1+z)\frac{{\rm d}q}{{\rm d}z}$, respectively. The negative values of deceleration parameter imply that the universe is expanding with an accelerating rate and values less than $-1$ indicate a super-exponential expansion. The jerk parameter on the other hand is a very useful parameter for investigating the deviation of a cosmological model from $\Lambda$CDM since its value stays pegged to unity in $\Lambda$CDM (ignoring the contributions other than $\Lambda$ and pressure-less matter), while it is in general a dynamical quantity in our model. In the case $(w_0,w_{\rm i})=(-1,-\frac{2}{3})$ the deceleration parameter evolves from $0.5$ as in the $\Lambda$CDM and goes ever monotonically to $-\frac{3}{2}$ which indicates a big rip end of the universe while the universe approaches asymptotically de Sitter universe with a deceleration parameter equal to $-1$ in $\Lambda$CDM model. We depict the evolution of $q$ and $j$ in redshift $z$ for $(w_0,w_{\rm i})=(-1,-\frac{2}{3})$ and $\Lambda$CDM in Fig \ref{fig:qj} using the values with $1\sigma$ errors from Table \ref{table:LVDPobs3}. We find that the current value of the deceleration parameter is $q_0=-0.545\pm0.014$ ($1\sigma$) and the universe starts its accelerating expansion at redshift value $z_{\rm tr}=0.647^{+0.021}_{-0.018}$ ($1\sigma$) while these values are $q_0=-0.547\pm0.012$ ($1\sigma$) and $z_{\rm tr}=0.665\pm0.021$ ($1\sigma$) in $\Lambda$CDM model. We observe that the jerk parameter exhibits a non-monotonic behavior but stays at values very close to unity, and find its current value is $j(z=0)=1.087\pm 0.001$ ($1\sigma$). Although the expansion history of the universe is almost the same in both models, they predict completely different futures. In the case $(w_0,w_{\rm i})=(-1,-\frac{2}{3})$ the universe enters into a super-acceleration regime ($q<-1$) at redshift $z_{\rm s}=-0.487^{-0.003}_{+0.004}$ ($1\sigma$) and ends with a big rip.

\section{Concluding remarks}

The new parametrization of DE we introduced in this paper allowed us to carry out successive observational analyses by decreasing its degrees of freedom systematically until ending up with a dynamical dark energy model having no additional parameters compared to $\Lambda$CDM. We first fixed the initial value of the EoS to some particular values and found some indications in favor of an evolving DE. We then additionally fixed the EoS parameter to $-1$ for the present epoch of the universe and found that EoS parameter that equals $-\frac{2}{3}$ at high redshifts and $-1$ at the present epoch gives a slightly better fit to the combined Planck+BAO+SN data than $\Lambda$. The variation of the EoS parameter in scale factor, on the other hand, is dynamical but always less than zero, which means that even today, when its EoS parameter is equal to $-1$, the DE does not genuinely mimic $\Lambda$. Thus we gave a cosmological model that can compete with $\Lambda$CDM but involving a dynamical DE. We also note that DE in this model passes below the phantom divide line in the future, hence leading to a big rip end of the universe. Accordingly, because the universe lives for finite time and the EoS parameter of DE evolves around $-1$, hence the energy density of DE is also dynamical, throughout the history of universe, our model relieves also the problems related with cosmological constant assumption of $\Lambda$CDM model.

\section*{Acknowledgments}
\"{O}.A. acknowledges the support by T\"{U}B{\.I}TAK Research Fellowship for Post-Doctoral Researchers (2218). \"{O}.A. and T.D. acknowledge the support from Ko\c{c} University. Authors thank for the hospitality of the Abdus Salam International Center for Theoretical Physics (ICTP), where part of this work was carried out.

\bigskip





\end{document}